\documentclass[12pt]{article}
\usepackage{epsfig}
\usepackage{epsf}
\newcommand{\bmat}{\left(\begin{array}}
\newcommand{\emat}{\end{array}\right)}

\def\yzero{\smash{\hbox{$y\kern-4pt\raise1pt\hbox{${}^\circ$}$}}}

\def\-{\hphantom{-}}

\def\s2{\frac{1}{\sqrt2}}

\def\beq{\begin{equation}}
\def\eeq{\end{equation}}
\def\beqa{\begin{eqnarray}}
\def\eeqa{\end{eqnarray}}

\def\IF{\relax{\rm I\kern-.18em F}}
\def\II{\relax{\rm I\kern-.18em I}}
\def\IP{\relax{\rm I\kern-.18em P}}
\def\IC{\relax\hbox{\kern.25em$\inbar\kern-.3em{\rm C}$}}
\def\IR{\relax{\rm I\kern-.18em R}}

\def\Dsl{\,\raise.15ex\hbox{/}\mkern-13.5mu D} 
\def\IZ{Z\kern-.4em  Z}
\def\bmat{\left(\begin{array}}
\def\emat{\end{array}\right)}

\def    \part          {\partial}
\def    \be            {\begin{equation}}
\def    \ee            {\end{equation}}
\def    \bea           {\begin{eqnarray}}
\def    \eea           {\end{eqnarray}}

\begin{document}
%
\makeatletter
\@addtoreset{equation}{section}
\makeatother
\renewcommand{\theequation}{\thesection.\arabic{equation}}
%
\pagestyle{empty}
\rightline{ULB-TH-98/18}
\vspace{0.5cm}
\begin{center}
\LARGE{Leptogenesis with virtual Majorana neutrinos\\[20mm]}
\large{J.-M. Fr\`ere, F.-S. Ling\footnote{Aspirant FNRS.}, 
        M. H.G. Tytgat\footnote{Collaborateur Scientifique FNRS.}
                and  V. Van Elewyck\footnote{Chercheur FRIA.}\\[8mm]}
\small{Service de Physique Th\'eorique, CP225\\[-0.3em]
                Universit\'e Libre de Bruxelles\\[-0.3em]
Bvd du Triomphe, 1050 Brussels, Belgium
\\[7mm]}
\small{\bf Abstract} 
\\[7mm]
\end{center}
\begin{center}
\begin{minipage}[h]{14.0cm}
We present  a mechanism of leptogenesis based on the out-of-equilibrium decay of 
a scalar particle into heavy virtual Majorana neutrinos. This scheme 
presents many conceptual advantages over the conventional scenario of Fukugita 
and Yanagida.
In particular, the standard techniques of quantum field theory can be used 
to compute the lepton asymmetry, without  resorting to   
the phenomenological approximations 
usually made  to describe unstable particles. 
This simplification allows us to address in a  
well-defined framework some issues raised  in the recent literature. 
We also show, in a toy model,  that a successful leptogenesis scenario 
is possible  and requires  a rather
light scalar particle,  $10^6 GeV < m <  10^{13} GeV$. 
A natural embedding of this scheme in a gauged 
unified theory encompassing the Majorana fermions seems however difficult.
\end{minipage}
\end{center}
\newpage

\pagestyle{plain}

%

\section{Introduction}

\label{sec:intro}

Leptogenesis is an attractive scenario for the origin of the  baryon number  of
the Universe. It rests on the idea that if an antilepton 
excess is created at a scale well above the electroweak phase transition, $T \gg 100
GeV$, it can be  very efficiently converted into a net  
baryon asymmetry by  spaleron-like  processes,
that violate $B+L$ but preserve $B-L$. In the simplest scenarios, 
the initial lepton 
asymmetry is created in the out-of-equilibrium decay of heavy Majorana
neutrinos~\cite{fuku}. 
Among other things, such a   scheme has the advantage of separating the step 
of $CP$ and $L$ violation from the step of $B$ violation, which occurs later 
through sphaleron-like  processes, and thus to avoid  the pitfalls of maintaining an 
out-of-equilibrium situation around the electroweak scale. (See {\em i.e.}~\cite{cohen}
for a review of electroweak baryogenesis scenarios.) Also, as the conversion of 
$L$ into $B$ takes place at equilibrium and is essentially complete, these
mechanisms are  
largely insensitive to the details of the non-perturbative baryon number
violating processes.  

As is well known, CP violation is a crucial ingredient of leptogenesis~\cite{sakharov}
and it here arises from the interference between 
tree-level  diagrams and  the absorptive part of one-loop diagrams. Traditionally, only
the one-loop vertex corrections  were
 taken into account in most calculations~\cite{kolb}, even though
 it was known~\cite{kuzmin} that 
the self-energy corrections, through which the different Majorana neutrinos
can mix, 
do also  contribute to the CP asymmetry.
In particular, in the framework of the wave-function formalism of Weisskopf and 
Wigner~\cite{wigner},  it has  been argued  that, in the limit of nearly 
degenerate Majorana neutrinos, the self-energy contribution could be 
significantly  larger than  the vertex term, thus giving an 
enhanced lepton asymmetry~\cite{bota,liu,paschos}. Using the exact solution of 
the wave equation with a complex matrix, this effect has been
 verified 
for  the case of two scalar 
flavours~\cite{rouletone}. The
wave-function 
approach is only a phenomenological approximation however, and one might wish for a 
more rigorous and systematic formulation. The problem, as is well-known, is
that unstable 
particles are outside the realm of conventional quantum field theory, as they cannot be 
asymptotic states of the S-matrix. In particular, the self-energy corrections cannot be 
absorbed into the field wave-function renormalization constant, without 
destroying the hermiticity of the lagrangian.\footnote{A more satisfying approach, that 
only slightly departs from the canonical rules of quantum field theory, has been 
advocated in~\cite{pilaftsis}.} Another, but related,  issue is that the
Majorana propagation eigenstates are not well-defined,
 an effect that leads to an ambiguity in
the initial conditions for leptogenesis.\footnote{This has been emphasized
in~\cite{rouletone} for instance, but   presumably is a well-known problem.} 

To address some of these problems, 
 it has been proposed in~\cite{plumone} to consider  lepton number
violating  scattering 
processes in which the Majorana particles  appear only in intermediate states,
like in 
\begin{equation}
\label{proc}
l_L \, \phi \rightarrow N^* \rightarrow l^c_L \, \phi^\dagger
\end{equation}
  where $l_L$ are the
left-handed Standard Model (SM) leptons, $\phi$ is  the Higgs doublet and  $N$ are
 off-shell (*)  Majorana neutrinos.   
Compared to the Majorana decay, the main advantage of considering processes like~(\ref{proc}) 
 is that  the rules
 of quantum field theory can be applied straightforwardly.  
It is  for instance manifest that the self-energy 
corrections to the propagator of the 
Majorana must be included at one-loop. What is less 
obvious is how much these corrections contribute  to the lepton
asymmetry. Actually, as has been shown 
in~\cite{roulettwo}, 
  unitarity implies that when all the scattering channels like~(\ref{proc}) are taken into
  account, the resulting lepton excess is precisely zero. This is actually
just the requirement of departure from equilibrium: as the initial and final
states in the processes~(\ref{proc}) are the {\em same}, at equilibrium no lepton
asymmetry can be created. Departure from equilibrium can  be provided by
the expansion of the Universe,  that effectively 
selects a subset of the  processes~(\ref{proc}) and can thus lead to
  a non-vanishing lepton 
asymmetry~\cite{roulettwo,paschos}. 

In the present paper, we  study  a different mechanism, that is
a variation on the scattering scenario of~\cite{plumone}. In section 2, we
will
 consider a heavy
scalar 
particle, $\chi$, that is allowed to decay  into 
light (unspecified but sterile)
right-handed  fermions and  heavy right-handed Majorana neutrinos. The Majorana
neutrinos  decay  into the left-handed SM leptons and  Higgs scalar. (See figure 1.) 
If we impose the mass 
hierarchy,
$$
m_M \gg m_\chi \gg m_\phi, m_l = 0,
$$
the   $\chi$ can be viewed as a  source of Majorana 
neutrinos. 
This scenario encompasses  the conceptual advantages of the scattering
processes~(\ref{proc})  but furthemore leads to the  production of a net lepton asymmetry. 
Among other things, we will  verify that the self-energy corrections do indeed
give a non-negligible contribution 
 to the asymmetry. We will study the limit 
of degenerate Majorana neutrino masses, and show that the asymmetry has
 a finite, well-defined
expression at one-loop. 
(Phase counting  shows that even in this case, CP violation effects are possible.)
Also, the asymmetry so obtained is directly
related to the initial abundance of the $\chi$, independently of the basis
chosen to define the Majorana states. 
We provide some  numerical calculation performed for two 
flavours that make the various contributions (vertex and self-energy)
to the asymmetry more explicit and 
give some useful estimates. Finally, we discuss 
the out-of-equilibrium conditions in the Early Universe, that puts limits on the 
 Majorana and $\chi$  masses. 
These constraints compel the 
leptogenesis scenario to involve a neutral scalar $\chi$ at a scale between $10^6 $ 
and $10^{13} GeV $.

In Section 3, we try to embed our scheme  in a more 
physically motivated framework. For definiteness, we have in mind 
a natural gauge extension of the model, for instance  $SO(10)$. For 
simplicity, we have confined our argument to its  subgroup 
 $SU(2)_L \times SU(2)_R \times U(1)$, which already imposes strong
constraints. As we will show, 
adding further gauge degrees of freedom has non-trivial consequences. In this 
framework, the $\chi^+$ is the singly  charged component of a  triplet of  
$SU(2)_R$ while the vacuum expectation value of the neutral component $\chi^0$ breaks
$SU(2)_R$ and gives a 
Majorana  mass to the right-handed neutrinos,
$$
\Delta_R 
\equiv
\left(
\begin{array}{cc}\chi^+ / \sqrt{2} & \chi^{++}\\ 
\chi^0 & -\chi^+ / \sqrt{2}\\  
\end{array}
\right).
$$
The  decay of the $\chi^+$ as the source of the lepton asymmetry
is however immediately ruled out: 
the annihilation  of  $\chi^+ \chi^-$ pairs into  photons 
 is much too fast, so that the charged $\chi$ stay in thermal equilibrium at the 
epoch of interest, $T \sim m_\chi$. The next possibility is to consider the decay of the 
neutral  $\chi^0$ (figure 8). Because the lepton number violating decay rate 
is relatively slow in this case, we  have to consider other competing rare decay 
processes. As  almost no dilution is allowed, an analysis of 
the dominant decay channels  reveals that   coupling of 
scalar particle to the $SU(2)_R$ gauge bosons is sufficient to totally damp out the lepton 
asymmetry. Adding more fields could resolve this
problem, but
at the price of simplicity.

\section{$\chi$ decay leptogenesis: self-energy and vertex 
corrections}
We  first 
concentrate on the various sources of CP violation  in the decay of the 
scalar particle $\chi$. The final state considered is a right-handed (sterile)
fermion accompanied by a left-handed lepton and a Higgs boson, and reached through 
the exchange of a virtual heavy Majorana particle $ \chi \rightarrow l_R 
N^* \rightarrow l_L l_R \phi $ (see diagram 1, figure 1). The Majorana neutrinos are 
labelled according to their mass; the following mass hierarchy guarantees that 
the intermediate Majorana neutrino is off mass-shell, 
$$
M_3 > M_2 > M_1 > m_{\chi} \gg m_l , m_{\phi}=0
$$
We want to study the particular consequences of this mass hierarchy on CP 
violation,  as a theoretical framework and as a possible realistic scenario for leptogenesis. 

The decay of the $\chi$  is 
expected to produce a lepton asymmetry, since the intermediate Majorana can 
couple to both lepton-antiboson and antilepton-boson, with  final 
total lepton number  2, 0 or -2 (figure
1).\footnote{This  assignment corresponds to  total lepton number,
  left-handed plus right-handed. If the right-handed fermions are sterile, only
  the left-handed lepton number matters for leptogenesis. Our conclusions are essentially
  independent of the charge assignement chosen.}
The most general Yukawa lagrangian for the particles involved in our 
scheme is of the form
\begin{equation}
\label{lag}
{\cal L}_{yuk} = g_{i j}\, {\overline L}_{Li} \Phi  R N_j + 
 G_{k l}\,\chi\, \overline {l^{c}}_{R k} 
  R N_l + h.c.
\end{equation}
where 
${L}_L$
stands for the left-handed SM leptons,
$\Phi$ is the Higgs doublet, $N$ are the heavy Majorana neutrinos, $l_R$ are the
light, sterile right-handed fermions,  and $R=\frac{1+\gamma_5}{2}$. 
Conventionnally, the Majorana states $N$ are chosen so that the Majorana mass 
matrix $M$ is real and diagonal, which is always 
possible.  The two Yukawa coupling matrices $g$ and $G$, however, cannot 
be diagonalised {\em simultaneously} with $M$, in general. 
The lagrangian~(\ref{lag}) leads to the  tree level decays for the $\chi$ of figure 1.
\begin{figure}[t]
\label{ftwoone}
\centerline{\epsfig{figure=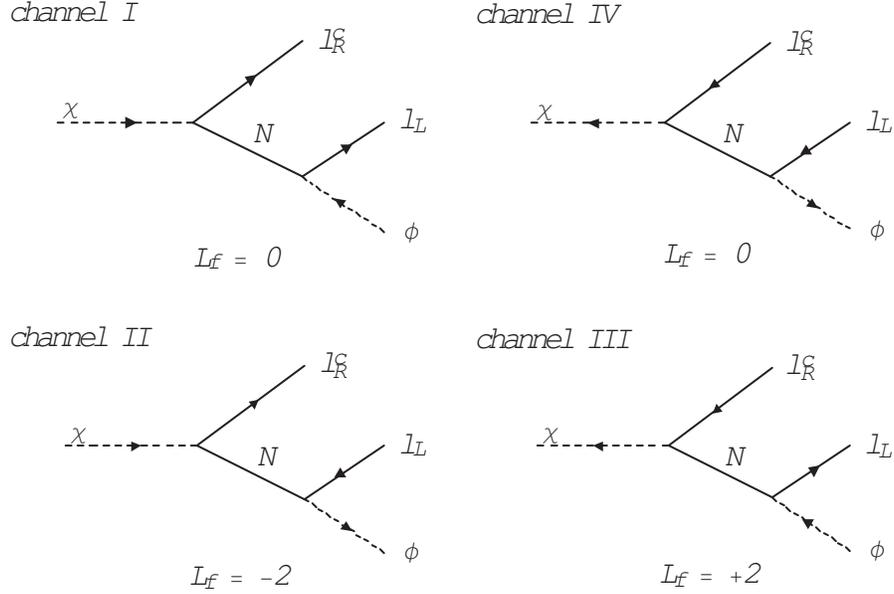}}
\caption{Decay channels of $\chi$}
\end{figure}
The decay channels form hermitian
conjugate pairs, {\em e.g.} II and
III, and CP violation becomes possible only at the one loop level, where 
tree level and one-loop diagrams can interfere. 
We expect that both loops including 
$\chi$ and $\phi$ scalars will contribute to the global asymmetry, since CP 
violating phases appear in the coupling matrices $g$ and $G$ in the most
general 
case. We can 
split the one-loop diagrams into self-energy loops and vertex corrections.
\begin{figure}[hbt]
\label{ftwotwo}
\epsfig{figure=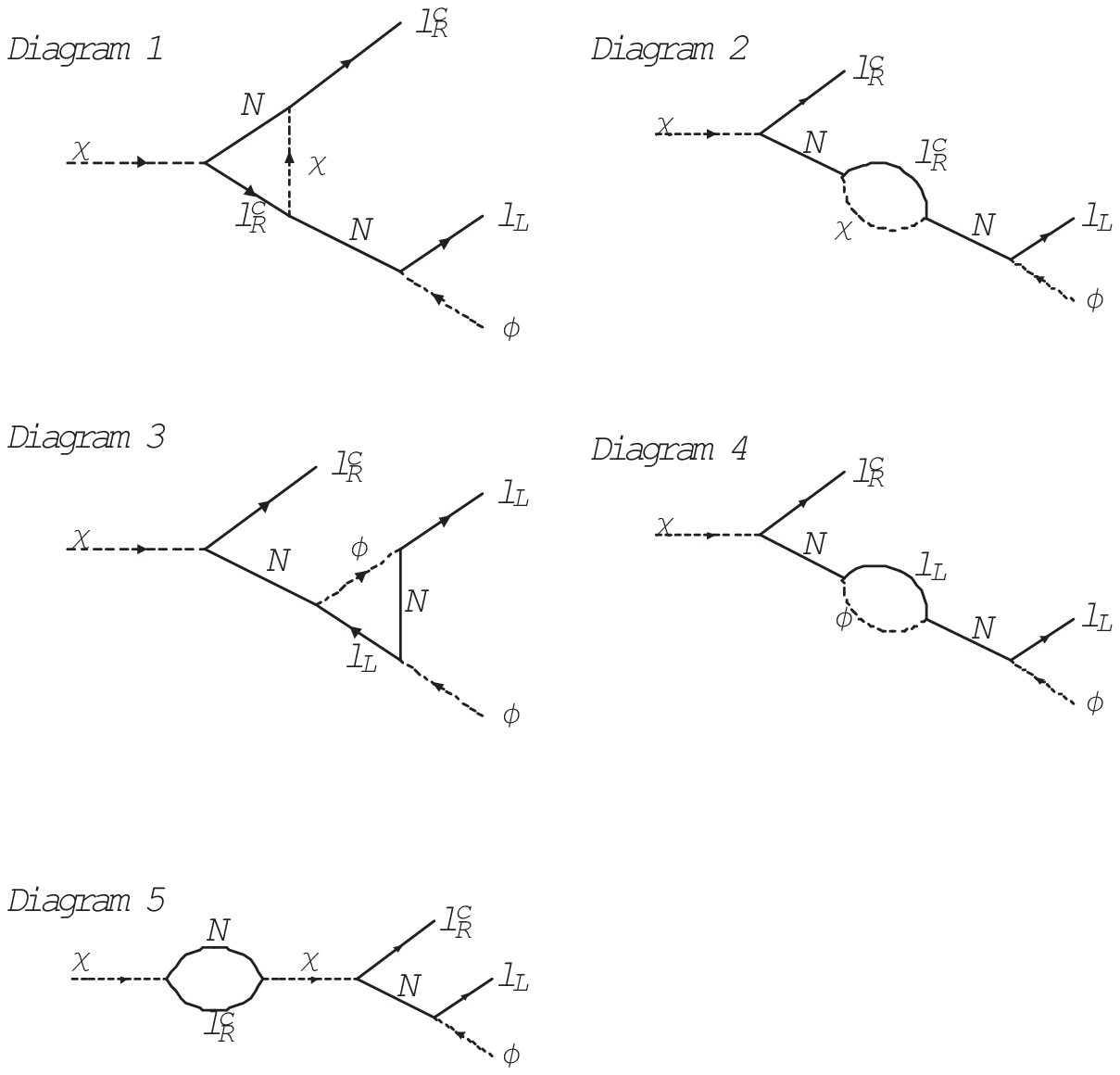}
\caption{One-loop corrections.}
\end{figure}
While a calculation including the vertex correction alone is well known to 
yield a non-vanishing asymmetry, the self-energy correction has been often 
neglected. However, it has been argued in~\cite{liu} and~\cite{paschos} that its 
contribution is
 far from negligible, and is even enhanced for nearly degenerate 
Majorana masses. 
In our scheme, the inclusion of self-energy loops is automatic 
since the Majorana neutrinos 
only appear as virtual intermediate states. 

According to the Cutkosky rules, the absorptive part of the one-loop diagrams, which 
provides the imaginary part 
needed for the CP asymmetry,  are given by the cut diagrams. 
Diagrams 1,2,5  of figure 2 are purely dispersive (no unitarity cut operates) 
and therefore don't contribute to the CP asymmetry.
In particular, the self-energy correction of  diagram 5 can 
be absorbed in the wave-function renormalisation of the $\chi$. Consequently, in this
scheme, the sources of CP violation are precisely the same as those relevant in the
conventional 
Majorana neutrinos decay scenario, namely, the asymmetry comes from loops 
involving the Higgs scalar. 

At tree level, the $\chi$ decay rate  is
\begin{equation}
\Gamma_0 [\chi \rightarrow l_{Rj} l_{Lk} \overline{\phi}] =
\frac {m_{\chi}}{64(2\pi)^3} 
\sum_{i,l} (g^{\dagger}g)_{li} (G^{\dagger}G)_{li} 
f^{(0)}(\frac {M_{i}}{m_{\chi}},\frac{M_{l}}{m_{\chi}})
\end{equation}
where
$$
f^{(0)}(x,y)=xy \left(1+\frac{(x^2-1)^2}{x^2-y^2} \ln (1-\frac{1}{x^2})
- \frac{(y^2-1)^2}{x^2-y^2} \ln (1-\frac{1}{y^2}) \right)
$$ 
This function has a well-defined limit  when the Majorana are degenerate,
$$
\lim_{y \rightarrow x} f^{(0)}(x,y)=2x^2-1+2x^2(x^2-1) \ln(1-\frac {1}{x^2}).
$$
while in the limit of very heavy Majorana neutrino $ m_{\chi} \ll M $,
\begin{equation}
\label{approxrate}
\Gamma_0 [\chi \rightarrow l_{Rj} l_{Lk} \overline{\phi} ]
\rightarrow
\frac { m^{3}_{\chi} }{3 \cdot 2^6(2 \pi)^3} \sum_{i,l} 
\frac{(g^{\dagger}g)_{li} (G^{\dagger}G)_{li}}{M_i M_l}
\end{equation}
At this order, the CP asymmetry splits into two parts that can be calculated 
separately. The vertex correction gives the following contribution to the asymmetry,
\begin{equation}
\epsilon_v = \frac 
  {\Gamma_{01,v}^{as.} [\chi \rightarrow l_{Rj} l_{Lk} \overline{\phi}]}
  {\Gamma_0 [\chi \rightarrow l_{Rj} l_{Lk} \overline{\phi}]}
\end{equation}
with
\begin{equation}
\Gamma_{01,v}^{as.} [\chi \rightarrow l_{Rj} l_{Lk} \overline{\phi}]=
\frac {m_{\chi}}{32(2\pi)^4} 
\sum_{i,l,n} \Im m (g^{\dagger}g)_{nl}(g^{\dagger}g)_{ni} 
(G^{\dagger}G)_{il} 
f_{v}(\frac {M_{i}}{m_{\chi}},\frac{M_{l}}{m_{\chi}},\frac{M_{n}}{m_{\chi}})
\end{equation}
The function $ f_{v}(x,y,z) $ is again  well-defined for degenerate 
Majorana masses. In the  limit $m_\chi \ll M$ as in Eq.~(\ref{approxrate}) above,
\begin{equation}
\label{vertexapprox}
\Gamma_{01,v}^{as.} [\chi \rightarrow l_{Rj} l_{Lk} \overline{\phi}]
\rightarrow 
\frac { m_{\chi}^{5} }{3 \cdot 2^{10}(2 \pi)^4} \sum_{i,l,n} 
\frac{\Im m ( (g^{\dagger}g)_{nl}(g^{\dagger}g)_{ni} (G^{\dagger}G)_{il})}
     {M^{2}_{i} M_l M_n}
\end{equation}
%
%
The self-energy correction gives
\begin{equation}
\label{asymself}
\epsilon_w = \frac 
  {\Gamma_{01,w}^{as.} [\chi \rightarrow l_{Rj} l_{Lk} \overline{\phi}]}
  {\Gamma_0 [\chi \rightarrow l_{Rj} l_{Lk} \overline{\phi}]}
\end{equation}
with
%
%
\begin{equation}
\Gamma_{01,w}^{as.} [\chi \rightarrow l_{Rj} l_{Lk} \overline{\phi}]=
\frac {m_{\chi}}{16(2\pi)^4} 
\sum_{i,l,n} \Im m (g^{\dagger}g)_{nl}(g^{\dagger}g)_{ni} 
(G^{\dagger}G)_{il} 
f_{w}(\frac {M_{i}}{m_{\chi}},\frac{M_{l}}{m_{\chi}},\frac{M_{n}}{m_{\chi}})
\end{equation}
%
%
The complete expression for $f_{w}$ is cumbersome, but 
it can be expressed in terms of simple functions. Again, the limit of 
degenerate Majorana masses gives a well defined value,
\begin{equation}
\label{selfdeg}
\lim_{M_i, M_l, M_n \rightarrow M} f_{w}
(\frac {M_{i}}{m_{\chi}},\frac{M_{l}}{m_{\chi}},\frac{M_{n}}{m_{\chi}})
 =  \frac {1}{16}- \frac {3M^2}{8m^{2}_{\chi}}+(\frac {M^2}{4 m^{2}_{\chi}}
+ \frac {3M^4}{8m^{4}_{\chi}}) \ln \frac {M^2- m^{2}_{\chi}}{M^2}
\end{equation}
Also, in the limit  of a light $\chi$, $ m_{\chi} \ll M $,
\begin{equation}
\label{selfapprox}
\Gamma_{01,w}^{as.} (\chi \rightarrow l_{Rj} l_{Lk} \overline{\phi})
\rightarrow
\frac { m_{\chi}^{5} }{3 \cdot 2^9(2 \pi)^4} \sum_{i,l,n} 
\frac{\Im m ( (g^{\dagger}g)_{ni} (g^{\dagger}g)_{nl} (G^{\dagger}G)_{il})}
     {M^{2}_{n} M_l M_i}
\end{equation}

In the conventional scenario,  with Majorana neutrinos in the initial state, 
there are unphysical singularities   $\propto 1/({M^{2}_{i}- 
M^{2}_{j}})$  in the expression of the self-energy term (see for
instance~\cite{rouletone} or~\cite{plumone}). They
 signal a breakdown of the perturbative
expansion for $\Delta M \sim \Gamma$, where $M$ and $\Gamma$ are
respectively the
mass and   width of the
Majorana neutrinos, and are responsible for the enhancement of the lepton
asymmetry
from the self-energy
contribution with respect to one from the vertex. 
The present scheme offers no such singularities, and the limit of degenerate
Majorana neutrinos is finite, Eq.~(\ref{selfdeg}). This    holds of course
 as long as the Majorana are off mass shell, which is guaranteed by our choice
 of mass hierarchy, because the mass of the $ \chi $ 
particle limits the  value of  $q^2$ flowing through the Majorana 
propagator to be at most $m_{\chi}^{2} \ll M_i^2$. In 
the case of 2-body scattering of light particles, the 
self-energy correction presents the same 
pole, because the centre of mass energy in a scattering process is not limited 
in principle, even if this scattering is supposed to occur in the thermal bath 
at the temperature $ T^2 \ll M^{2}_{i} $, i.e. at the epoch when the lightest 
heavy Majorana neutrino decays. 
%
%
%
This pole enhances the contribution of the  self-energy compared to the
one from the vertex in the degenerate limit, but necessitates one to use
either a quite elaborate resummation scheme (for instance as in 
~\cite{plumone} which is limited to weak mixing) or a wave-functional 
approach~\cite{rouletone}.

The absence of such singularities is another non-negligeable 
simplification  offered by the present approach. On the other hand, there is 
no large enhancement $\sim
M/\Gamma$ either (provided $m_{\chi} < M $).
However,  the asymmetry is well defined for all  
Majorana mass patterns, including the limit  of degenerate Majorana masses, for
which the asymmetry does not necessarily vanish. 

Let us also re-emphasise that,  contrarily to the scattering
 processes of~(\ref{proc}) studied
in~\cite{plumone,roulettwo}, 
a true lepton asymmetry can be produced here. This is 
simply because the $\chi$ is unstable below $T <
m_\chi$ while the inverse decay, 
or recombinations is Boltzmann suppressed, which is another way to 
state that the unitarity 
constraint does not apply. Turning now to  leptogenesis, the framework of this 
mechanism is just the basic out-of-equilibrium decay of unstable particles. The 
various rates and CP asymmetries can be computed without reference to the 
Majorana propagation eigenstates. We only have to ensure that the conditions of an 
out-of-equilibrium decay for the $\chi$ are verified.

We have performed an explicit (numerical) calculation of the  ratio 
between the self-energy and the vertex contributions to the asymmetry, in the 
special case of two flavours. In the restricted parameter area spanned 
by the two lightest 
Majorana neutrinos, the matrices appearing in the  expressions of the
asymmetries can be 
parameterised as follows: the Majorana mass matrix can always be set to a real 
positive diagonal form
$$
M = 
\left(
\begin{array}{cc}
M_1 & 0\\
0 & M_2\\
\end{array}
\right)
$$
while 
$g$ is a complex matrix.  As the product $gg^{\dagger}$ can be diagonalized
by a unitary transformation on the left-handed leptons, we can  
 parameterise $g$ as
$$
g = 
\left(
\begin{array}{cc}
g_1 e^{i \alpha} & - \frac {g_{0}^{2}}{g_2} e^{-i \beta}\\
\frac {g_{0}^{2}}{g_1} e^{i \beta} & g_2 e^{-i \alpha}\\
\end{array}
\right),
$$
There is a similar parameterisation for $G$. 
Already with two flavours, there are four CP violating phases, even in 
the degenerate case, because the  two Yukawa coupling matrices limit the redefinition 
freedom on the right-handed neutrinos. 
This has to be contrasted with the conventional 
scenario, where only the two matrices $M$ and $g$ are present, so that an 
appropriate unitary rotation $U_R$ can eliminate the two CP violating phases of 
$g$ in the degenerate case. In the present case, these two phases would  only
be moved 
from $g$ to $G$. Of course, if $G$ is proportional to $g$, $M$ or the
identity matrix, 
both the vertex and self-energy contributions to the asymmetry vanish. (As  
shown in ~\cite{branco}, even with the matrices $M$ and $g$ alone, one CP 
violation phase can remain in the mass-degenerate limit if all three flavours 
are taken into account.) 
Figure 3 presents the ratio $ r = \epsilon_w / \epsilon_v $ plotted 
against the mass hierarchy $ h = M_2 / M_1 $, in the case where
the $\chi$ particle mostly couples to the 
lighter Majorana neutrino.
\begin{figure}[htb]
\label{ftwothree}
\centerline{\epsfig{figure=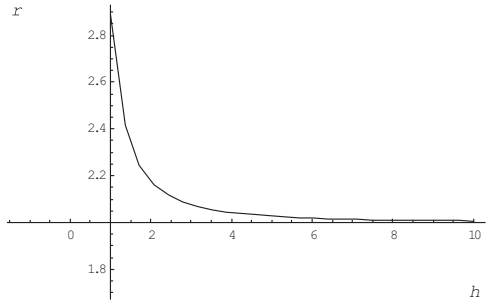,height=4cm}}
\caption{Ratio $r = \epsilon_w/\epsilon_v$ {\em vs} mass ratio $h = M_1/M_2$.}
\end{figure}
For large mass splitting, {\em i.e.}  when $ h \gg 1 $,  the 
self-energy term becomes twice  the vertex one, $ r \rightarrow 2 $ .
This limit was previously obtained
in~\cite{rouletone} for scalar  neutrinos 
decays. 
%
%
It is always true in the case $ m_\chi \ll M $, as  can be verified 
in the equations ~(\ref{vertexapprox}) and ~(\ref{selfapprox}).
But in the general case, the asymptotic value of this ratio depends on
the coupling matrices elements. 
On the other hand, in  the degenerate case,  
the self-energy contribution doesn't generally vanish, and can even 
have its maximum value at this point. The shape of the ratio $r$ against 
$h$ can be very different according to the choice  of  the coupling matrix 
$G$,
but generally, the self-energy contribution to the CP asymmetry is
 non-vanishing, 
even in the mass-degenerate limit. 
Also, as already stated, to consider the Majorana neutrinos  as 
intermediate 
states only permits to circumvent the problem of defining the  propagation eigenstates.
The current attempts to define the external states 
(see~\cite{rouletone} and~\cite{plumone} for example) are based on some 
diagonalisation which necessarily uses non-unitary transformations so that the 
resulting propagation eigenstates don't have a physical 
significance.

Before going on, it might be of interest to see why the cancellations met in 
the 2-body scattering scenario of~\cite{plumone} do not occur  here. 
This can be seen from the diagrammatic representation of "cut blobs", 
following~\cite{roulettwo}. To be concrete, we have chosen
$$
G = 
\left(
\begin{array}{cc}
1 & 0\\
0 & 0\\
\end{array}
\right)
$$
\begin{figure}[hbt]
\label{figtwofourone}
\centerline{\epsfig{figure=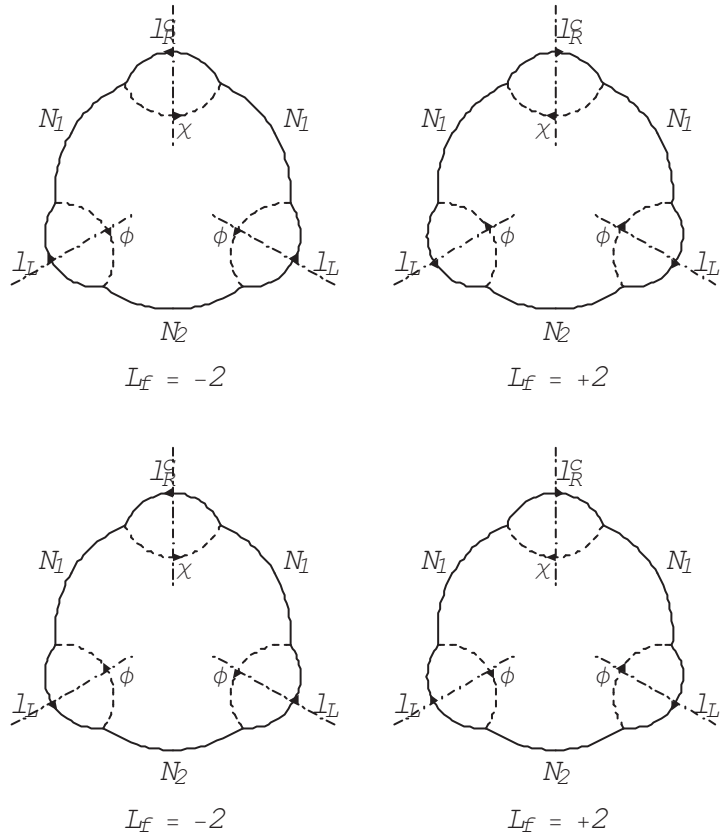}}
\caption{Interference diagrams with $\vert  L_f\vert =2$}
\end{figure}
{\em i.e.}  $\chi$  is only coupled to $N_1$. In figures 4  
and 5, we have only drawn  
the diagrams for which the cancellation with conjugates\footnote{The 
blobs of the second columns are the conjugate 
of those of the first columns. Note that the left-handed lepton current in the
self-energy
correction flips sign between
the
first and second row.} is not immediate. 
\begin{figure}[hbt]
\label{figtwofourtwo}
\centerline{\epsfig{figure=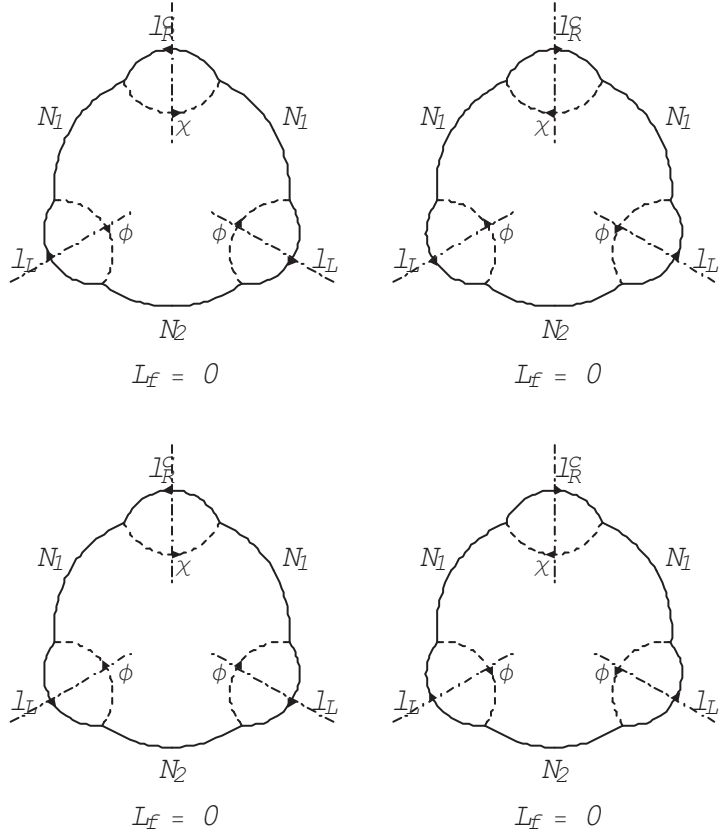}}
\caption{Interference diagrams with $ L_f = 0$}
\end{figure}
The one-loop self-energy 
diagram consists of two parts, one with the left-handed lepton and Higgs scalar running one 
way, and the other with the arrows reversed. Then, the diagram with the lepton
in 
the self-energy loop 
running from $N_2$ to $N_1$ cancels with its CP conjugate, while the other diagram
 must be added to the one with  $L_f = 0$ to find a 
complete cancellation. Hence, only  when all the decay channels of the $\chi$ particle 
are taken into account,  is there a full cancellation of the various (unweighted) 
CP self-energy asymmetries, but the lepton asymmetry produced by the self-energy 
in the decay channel does not vanish. This reasoning shows that, as expected, 
the lepton asymmetry is 
produced through the interference between $ L = 0 $ and $ L \neq 0 $ channels, and that
CP violation is observed only when specific channels (namely those with $ L \neq 0$
are selected. Finally, like in the scattering processes ~(\ref{proc}),
 a full cancellation of the lepton asymmetry occurs
if  the inverse decay channels are taken into account~\cite{roulettwo}. In the early universe,
 these channels are Boltzmann  suppressed if the $\chi$ decays out-of-equilibrium.

The last question is to see whether our scheme can yield enough lepton
asymmetry in the early universe. At this point, we must impose that the $ \chi $ is a
neutral particle, and for simplicity, we will assume a single real scalar,
 that couples to heavy Majorana neutrinos and some
light,  {\em sterile} right-handed fermions. 
Indeed, if the  $\chi$ particles were charged, they would pair-annihilate
very efficently into photons at $T \sim m_\chi$. The resulting lepton
asymmetry
  would be
Boltzman suppressed and too small for practical purposes.

From our  calculations, a lepton  excess $n_L=L/s \sim  10^{-10}$,
where $s$ is the entropy density of the universe at the epoch of interest,
 can be reached  for  $ x=m_{\chi}/M > 
10^{-2} $. Figure 6 shows for instance a plot of the two contributions, vertex
and self-energy, to the asymmetry in the 
case of degenerate Majorana mass, which gives the largest  asymmetry for a fixed scalar 
mass.
\begin{figure}[hbt]
\label{ftwofive}
\centerline{\epsfig{figure=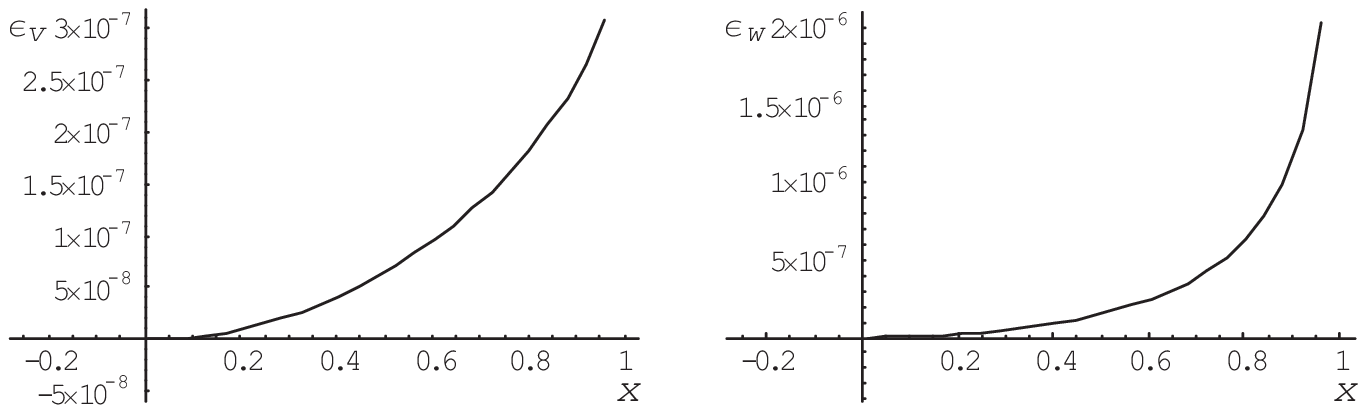}}
\caption{Asymmetry $ \vert \epsilon_v \vert $ and 
         $ \vert \epsilon_w \vert $ versus $x=m_{\chi}/M$}
\end{figure}
As the value of the asymmetry increases when the scalar mass approaches 
the Majorana mass,  it seems that it would suffice to take these  as 
close as possible to maximise the lepton asymmetry. For 
 leptogenesis, this has however a limitation. Suppose that the 
temperature is about the mass of the lightest Majorana neutrino $N_1$ 
and that 
$N_1$ 
is out-of-equilibrium. Then, since the 2-body lepton number violating 
scattering $l_L \overline{\phi} \leftrightarrow \overline{l_L} \phi $ is also 
out-of-equilibrium, the decay of the Majorana neutrinos would also produce a net 
lepton asymmetry. In the thermal bath, this process will compete with the scalar 
decay. In that case, an estimate for the asymmetry must take the coupled 
evolution of both Majorana neutrinos and scalars into account.

For
definiteness, we have chosen to consider a 
mechanism of leptogenesis where $L$ 
is evaluated from $\chi$ decays 
alone.  We thus  impose the following conditions,
\begin{eqnarray}
\label{cone}
\alpha_g M > g_{*}^{1/2} \frac {M^2}{M_{Planck}} \\
\label{ctwo}
\alpha_{g}^{2} \frac {m_{\chi}^{3}}{M^2} < g_{*}^{1/2} \frac { m_{\chi}^{2}} 
{M_{Planck}} \\
\label{cthree}
\alpha_g \alpha_G \frac {m_{\chi}^{3}}{M^2} < g_{*}^{1/2} \frac { m_{\chi}^{2}} 
{M_{Planck}} \\
\label{cfour}
r= \frac {m_{\chi}}{M} > 10^{-2}
\end{eqnarray}
Equation~(\ref{cone}) guarantees that the  lightest Majorana $N_1$ is {\em in} thermal
equilibrium when $T \sim M_1$, so that its decay does not 
generate any lepton asymmetry. Equation~(\ref{ctwo}) ensures that all 2-body 
lepton number violating processes are out-of-equilibrium when the $\chi$  start to 
decay. Equation~(\ref{cthree}) is the requirement that $\chi$ decays
 out-of-equilibrium. Lastly,~(\ref{cfour}) is needed to have enough 
 lepton asymmetry, . 

These four constraints limit the  mass of the Majorana  and $\chi$ 
 to be within the domain {\em D} depicted in figure 7.
 (We have also imposed $ m_{\chi} < M_M$ and we have taken $\alpha_g
 \equiv g^2/4 \pi$ 
and $\alpha_G \equiv G^2/4 \pi \sim 10^{-4}$ 
while $g_* \sim 10^2$ counts the number of degrees of freedom at the epoch of interest.)
\begin{figure}[hbt]
\label{ftwosix}
\centerline{\epsfig{figure=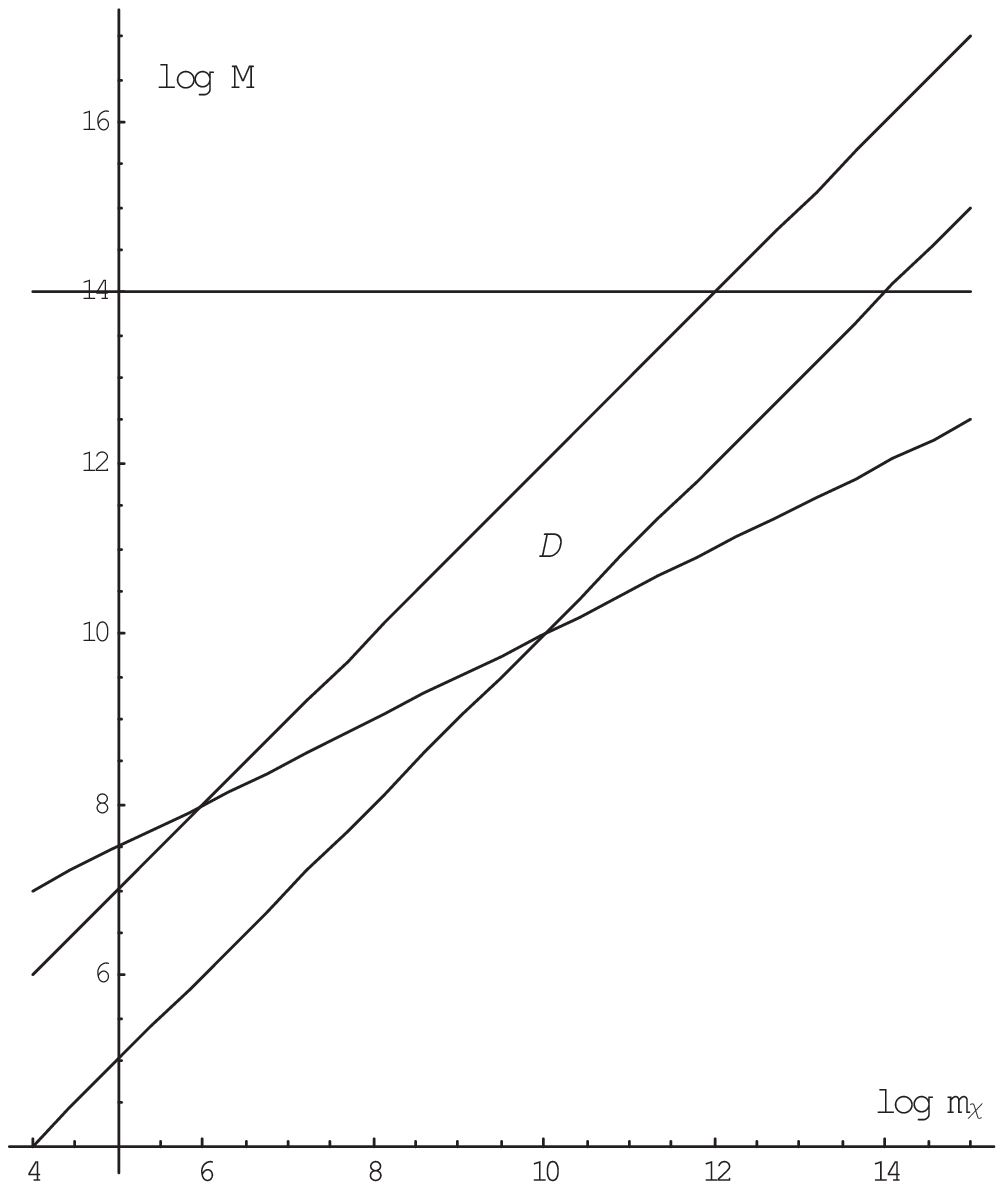}}
\caption{Mass scales constraints in log-log plot}
\end{figure}
The allowed  domain is quite large and not very 
constraining for the values of $ m_{\chi}$.
The constraint from $n_L \sim  10^{-10} $ gives a  lower bound on the scalar mass, 
$ m_\chi > \sim 10^6 GeV$. In the conventional Majorana decay  scenario there is a  lower
bound on the Majorana mass imposed  by the condition that the neutrino  
decays out-of-equilibrium;  in contrast, because we consider a three-body
decay process which has a different energy dependence, the same condition 
gives here an upper bound on the mass of the $\chi$, 
$m_\chi \ll 10^{13} GeV$. 

From the discussion of the present section, it appears that our toy model 
provides a {\em consistent} mechanism of leptogenesis.
The existence of a scalar coupled to Majorana neutrinos is however  
 suggestive of an extension of the SM to $SO(10)$ or at least its
 subgroup  $G = SU(2)_L \times SU(2)_R \times U(1)$ for instance.
We analyze  in section 3 how our model would fare in such a scheme.

\section{Effects of dilution in the gauged version of the  model}

The simplest way to embed the $ \chi$
 particle into the framework of a gauge 
theory is to work in a left-right model, i.e. with the group $ SU(2)_L \times
 SU(2)_R 
\times U(1)$.
 The scalar sector contains (at least) one complex scalar 
$\Delta_R \sim (0,1,2)$ which is in a (complex) triplet representation of
 $SU(2)_R$,
 similar scalars that couple to $SU(2)_L$,  $ \Delta_L \sim
 (1,0,2)$ and a  Higgs bi-doublet $\Phi \sim 
(1/2,1/2,0)$. 
$$\Delta_R \equiv\left(\begin{array}{cc}\chi^+ / \sqrt{2} & \chi^{++}\\ 
\chi^0 & -\chi^+ / \sqrt{2}\\  
\end{array}
\right)
$$
The Yukawa coupling of $ \Delta_R $ to the right-handed lepton doublet $ L_R $ 
is given by
$$
{\cal L}=-G \overline{L_{R}^{c}} \Delta_R L_R + h.c.
$$
The charged component $ \chi^+ $ and the neutral component $ \chi^0 $ couple to 
the right-handed neutrinos while the doubly charged $\chi^{++} $ only couples to 
charged leptons. In this scheme, 
the vacuum expectation value $\langle \chi^0 \rangle = V_R$ of the neutral component gives 
the Majorana mass term to the right-handed neutrinos (we will take it to br real).
 As it is  possible to 
choose a basis in which the Majorana mass matrix is real diagonal, 
the coupling of the right-handed lepton doublet to 
the scalar triplet is then also real diagonal (in the case of one triplet).
$$
G=\frac{1}{V_R}
\left(\begin{array}{ccc}M_1 & 0 & 0\\ 
0 & M_2 & 0\\  
0 & 0 & M_3\\  
\end{array}
\right)
$$
In the broken phase, the remaining gauge group is $ SU(2)_L \times U(1)$. The  real, massive  $ \chi^0 $ couples to two 
Majorana neutrinos, and the $\chi^+$ ( the $ \Im m \chi^0 $ ) component is  the longitudinal
component of the massive  gauge boson $ W_R $ ($Z_R$).
As the $ W_R $ (or the Goldstone $ \chi^+ 
$) are charged particles, they can annihilate into photons,  and this process occurs 
much faster than the decay into light leptons, so 
that, as already stated in section 2, these particles are not suitable for leptogenesis.
$Z_R$, being a heavy gauge boson, also abondantly decays into fermion pairs,
and its decay leads to inacceptable dilution of the CP asymmetry in this channel.

If we want  leptogenesis to occur in this {\em minimal}  framework,
we have to turn to the decay modes of the $\chi^0$,  that occurs through two
intermediate 
Majorana neutrinos, and yields  four light particles in the final state (see figure 8).
$$
{\cal L}_{\chi}=G \overline{N} \chi N
$$
\begin{figure}[hbt]
\label{fthreeone}
\centerline{\epsfig{figure=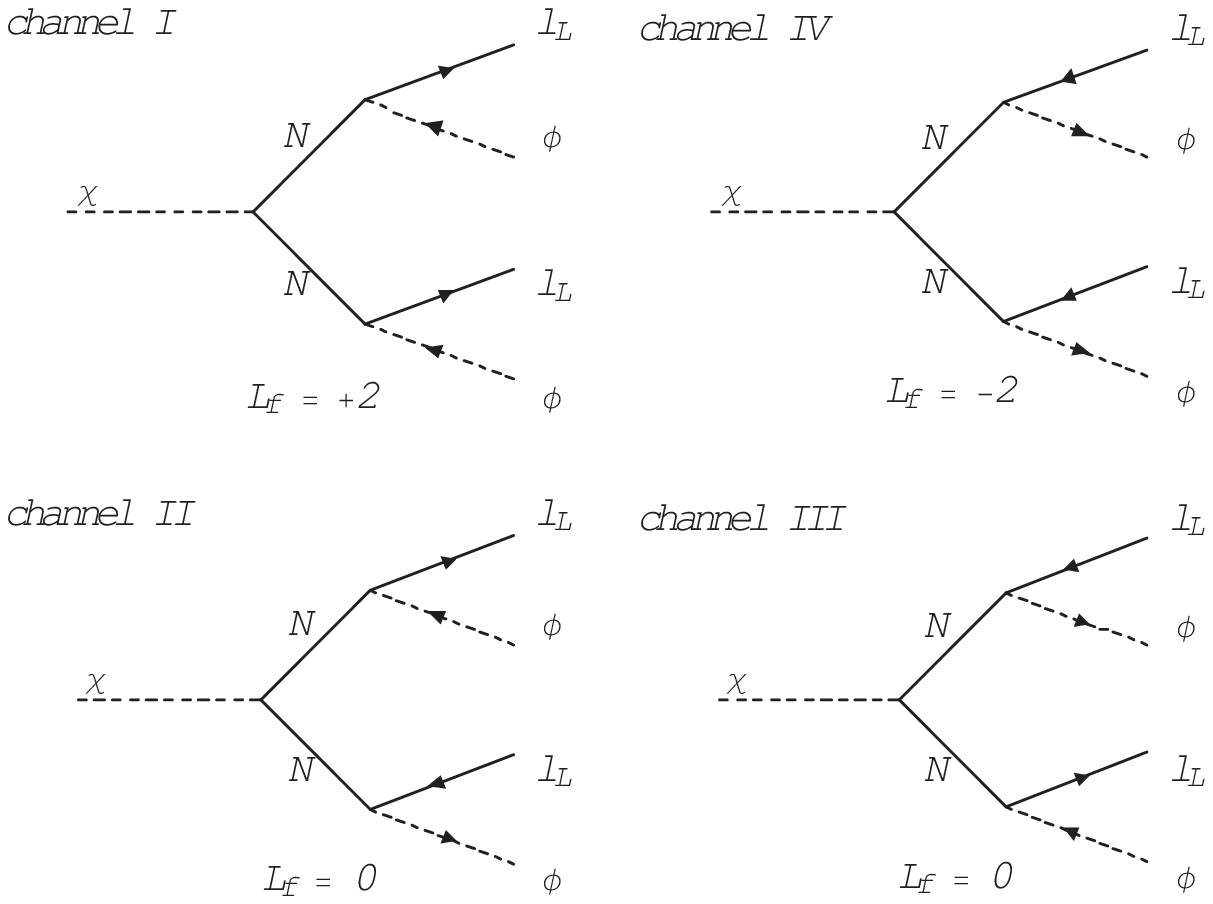}}
\caption{4-body decays of the neutral $\chi$ in the $LR$ symmetric model}
\end{figure}
The problem we have to face now is that a four-body decay rate 
is generally quite small. We  estimate that
$$
\Gamma_0 (\chi \rightarrow ll \overline{\phi} \overline{\phi}) \sim 10^{-7}  
\sum_{l,j} (g^{\dagger}g)^{2}_{lj} 
\frac {m_{\chi}^{5}}{M_j M_l V_{R}^{2}}
$$
The lepton asymmetry produced in this case is of the same 
order as the one  obtained for the  three-body decay discussed in section 2.
Consequently, only 
a small dilution of the lepton asymmetry by the other, lepton-number conserving, 
decay channels of the $\chi^0$ is acceptable. 

For one thing, this requires some fine tuning in the Higgs sector to prevent the $\chi^0$ to
decay into other light scalars. But much more problematic are the
 other rare decay channels of the $\chi^0$ that can occur at
 one-loop, or through higher dimensional operators.
%
%
leptons (figure 10).
%
%
Consider for instance  the one-loop decay mode of the $\chi^0$ into two 
photons 
(or two gauge bosons $a^{\mu}$ associated with $U(1)_Y$) as in
figure 9.
\begin{figure}[hbt]
\label{fthreefour}
\centerline{\epsfig{figure=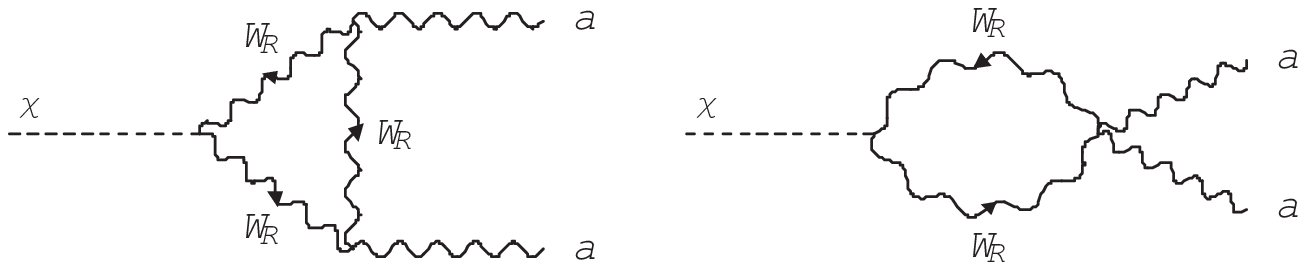}}
\caption{decay of $\chi$ to massless gauge bosons at one-loop level}
\end{figure}
This case is  similar to the decay of the Higgs  into photons in the SM, 
which has been computed in~\cite{ellis};it would give
$$
\Gamma_{\chi \rightarrow \gamma \gamma} \sim 10^{-5} g_R^{6} \frac 
{m_{\chi}^{3}}{M_{W}^{2}}
$$
$g_R$ stands for the coupling constant of the right-handed gauge bosons. This rate
is much 
larger than our
estimate of $ \Gamma_0 (\chi \rightarrow ll \overline{\phi} \overline{\phi}) $
if $g_R \sim g$, and the two-photons decay mode gives the dominant dilution factor of the lepton 
asymmetry. 

This constraint arises only because the scalar is related to 
 the same symmetry breaking scale as the 
corresponding massive gauge bosons.  In principle, it should be
possible to construct a model with a light neutral
scalar that can only decay into light sterile right-handed fermions and heavy
Majorana neutrinos, for instance by adding more scalar representations, but
as this detracts from the simple gauge structure of the model, we will not pursue this here.

\section*{Conclusions}

In a search for a clarification of the leptogenesis scheme, 
we have analysed a situation in which the Majorana neutrinos
appear only as intermediate states. 
We have  computed the lepton asymmetry without need to worry 
about defining propagation eigenstates for the unstable Majorana particles
while including  the CP violating effect from both the vertex and self-energy 
one-loop corrections. We have shown in section 2 that this very simple, 
but accordingly
{\em ad hoc} scheme  provides not only a
consistent but also efficient mechanism of leptogenesis. 
The embedding  in a realistic minimal gauged extension of the SM  is however 
problematic.  In particular, the addition of gauge degrees of freedom 
 gives  more dilution of the lepton asymmetry than is admissible. 
We have not considered more elaborate phenomenological models because 
 our main interest was to address a question of
principle --how to include all the CP violationg effects  in a self-consistent
an systematic framework. Obviously, a more general framework, that would
allow to compute  CP violation effects with Majorana neutrinos starting from
first 
principles, and including the influence of a medium (like in the Early 
Universe) would
be most welcome.

\section*{Acknowledgments}

This work was partially supported by the IISN (Belgium), and by the 
Communaut\'e 
Fran\c caise de Belgique - Direction de la Recherche Scientifique 
programme ARC.

\end{document}